\documentclass[
 amsmath,amssymb,
 aps,
prd,final,twocolumn,letterpaper,nofootinbib,superscriptaddress]{revtex4}

\usepackage{tabu}
\usepackage{graphicx}
\usepackage{dcolumn}
\usepackage{bm}
\usepackage{slashed}
\usepackage{cancel}

\usepackage{caption}
\usepackage{enumerate}
\usepackage{verbatim}
\usepackage{dsfont}
\usepackage{bbold}

\usepackage{color}
\usepackage{xcolor}
\usepackage[hyperfootnotes=false]{hyperref}
\usepackage{amsmath,braket}
\hypersetup{colorlinks=true, linkcolor=magenta, filecolor=cyan, urlcolor=blue}

\def\@Aboxed#1&#2\ENDDNE{%
  \settowidth\@tempdima{$\displaystyle#1{}$}%
  \addtolength\@tempdima{\fboxsep}%
  \addtolength\@tempdima{\fboxrule}%
  \global\@tempdima=\@tempdima
  \kern\@tempdima
  &
  \kern-\@tempdima
  \fcolorbox{red}{yellow}{$\displaystyle #1#2$}
}

\newlength\dlf

\newcommand{\be}{\begin{equation}}
\newcommand{\ee}{\end{equation}}
\newcommand{\bea}{\begin{eqnarray}}
\newcommand{\eea}{\end{eqnarray}}
\newcommand{\bma}{\left(\begin{array}}
\newcommand{\ema}{\end{array}\right)}

\newcommand{\intd}{\mathrm{d}}

\usepackage{tikz-feynman}
\tikzfeynmanset{compat=1.1.0}
\tikzset{graviton/.style={decorate, decoration={snake, amplitude=.4mm, segment length=1.5mm, pre length=.5mm, post length=.5mm}, double}}
\begin{document}

\preprint{APS/123-QED}

\title{Higgs Inflation and the Electroweak Gauge Sector}
\author{Stephon Alexander}
\email{stephon\_alexander@brown.edu}
\affiliation{%
Brown Center for Theoretical Physics \\
Department of Physics, Brown University, Providence, Rhode Island 02912, USA}%

\author{Cyril Creque-Sarbinowski}
 \email{ccreque@flatironinstitute.org}
 \affiliation{%
Center for Computational Astrophysics
\\Flatiron Institute, 162 5th Avenue, New York, New York 10010, USA}
 
\author{Humberto Gilmer}
\email{humberto\_gilmer@brown.edu}
\affiliation{%
Brown Center for Theoretical Physics \\
Department of Physics, Brown University, Providence, Rhode Island 02912, USA}%

 \author{Katherine Freese}
 \email{ktfreese@utexas.edu}
 \affiliation{Texas Center for Cosmology and Astroparticle Physics, Department of Physics, The University of Texas at Austin, Austin, Texas 78712, USA}

\date{\today}
             
\begin{abstract}
We introduce a new method that allows for the Higgs to be the inflaton. That is, we let the Higgs be a pseudo-Nambu-Goldstone (pNG) boson of a global coset symmetry $G/H$ that spontaneously breaks at an energy scale $\sim 4\pi f$ and give it a suitable $SU(2) \subset G$ Chern-Simons interaction, with  $\beta$ the dimensionless Chern-Simons coupling strength and $f$ an $SU(2)$ decay constant. As a result, slow-roll inflation occurs via $SU(2)$-induced friction down a steep sinusoidal potential. In order to obey electroweak  $SU(2)_{\rm L}\times U(1)_Y$ symmetry, the lowest-order Chern-Simons interaction is required to be quadratic in the Higgs with coupling strength $\propto \beta^2/f^2$. Higher-order interaction terms keep the full Lagrangian nearly invariant under the approximate pNG shift symmetry. Employing the simplest symmetry coset $SU(5)/SO(5)$, $N$ $e$-folds of inflation occur when $N \approx 60 \left(g/0.64\right)^2\left[\beta/\left(3\times 10^6\right)\right]^{8/3}\left[f/\left(5\times 10^{11}\ {\rm GeV}\right)\right]^{2/3}$, with $g$ the weak isospin gauge coupling constant. Small values of the decay constant, $f \lesssim 5 \times 10^{11} {\rm GeV}$, which are needed to address the Higgs hierarchy problem, are ruled out by electric dipole measurements and so successfully explaining inflation requires large $\beta$. We discuss possible methods to achieve such large couplings and other alternative Higgs inflation scenarios outside the standard modified-gravity framework. 
\end{abstract}
\maketitle
\section{\label{sec:intro}Introduction}
Inflation posits the existence of a slowly-rolling scalar field, the inflaton, to drive a period of accelerated expansion in the early Universe as a means to solve several problems in cosmology~\cite{Guth:1980zm}. Given that we already have a scalar field in the Standard Model (SM), the Higgs~\cite{Englert:1964et, Higgs:1964pj, Guralnik:1964eu}, a possible minimal realization of inflation identifies the inflaton as the SM Higgs.\footnote{Other possible SM minimal realizations include those that identify the inflaton as a neutrino condensate~\cite{0811.2998}} The SM Higgs potential, however, is not flat enough to both reproduce the observed matter power spectrum and sustain a sufficiently long inflationary phase, as its quartic self-interaction is too large~\cite{Adams:1990pn, 1307.0708, 1807.02376, Linde:1981mu, Linde:1983gd}. As a result, scenarios beyond the SM must be considered to achieve successful Higgs inflation. 

The original method to induce a slowly-rolling Higgs invoked a non-minimal gravitational coupling that exponentially flattened the SM Higgs potential upon a field redefinition~\cite{Salopek:1988qh, Bezrukov:2007ep}. Since its original formulation, a variety of extensions and generalizations, such as those involving derivative-Higgs couplings~\cite{1003.2635, 1003.4285, 1008.4457, 1012.4238, 1203.4059}, alternative formulations of General Relativity~\cite{0803.2664, 1012.2900, 1709.07853, 1811.09514, 1910.03488, 1901.01794, 1904.05699, 2209.11051}, warm initial conditions~\cite{2105.14552, 2303.00572}, additional particles~\cite{0809.3395, 0912.2718, 1107.2163, 1604.06760, 1405.7331, 1606.02202, 2207.05484}, and other approaches involving SUSY or other non-minimal couplings ~\cite{Yin:2022fgo, Choudhury:2013zna, Choudhury:2017cos} have been explored, with all such varieties flattening the Higgs potential in a similar manner. The original non-minimal proposal also violates perturbative unitarity due to the small EFT scale of the coupling compared to the high-energy scales of inflation~\cite{0902.4465, 0903.0355, 1002.2730, 1002.2995}. As a result, numerous perturbative and non-perturbative unitarity resolutions inside~\cite{0912.5463, 1005.2978, 1008.5157, 1010.1417, 1011.4179, 1112.0954, 1301.1787, 1306.6931, 1501.02231} and outside~\cite{1711.08761, 2002.07105, 2007.04111, 2110.03925, 2111.05621} the original framework have been proposed, with the predictiveness of non-minimal Higgs inflation depending on the UV completion considered~\cite{1402.1476}. Given this predictive uncertainty, it is thus desirable to further investigate a variety of alternative Higgs inflation models. The scope of all alternatives, however, is not infinite, as one must be able to also solve the eta problem, whereby higher-order corrections to the inflaton potential spoil its desired flatness~\cite{hep-ph/0404012}.

One paradigm meant to address the eta problem is natural inflation, where  the inflaton is endowed with an approximate shift symmetry that protects the shape of its potential~\cite{Freese:1990rb}. In natural inflation, the inflaton is a pseudo-Nambu-Goldstone (pNG) boson with a sinusoidal potential derived from the UV physics of a strongly-interacting vacuum. Cosmic microwave background observations have pushed the original variant of this model to have large super-Planckian excursions~\cite{2110.00483}, but its general form remains attractive~\cite{Arkani-Hamed:2003xts, Arkani-Hamed:2003wrq, Freese:2004un, Ben-Dayan:2014lca, Grimm:2014vva, Choi:2014rja, Czerny:2014wza, Kappl:2015esy, Linde:2018hmx, Freese:2021noj}. In particular, if the inflaton is able to efficiently dissipate its kinetic energy through friction, then sub-Planckian excursions, characterized by a steep potential, do not pose an issue. Such a solution was first proposed through the use of Abelian gauge fields~\cite{0908.4089,aps} and then followed up with its non-Abelian variety ~\cite{Maleknejad:2011jw, Maleknejad:2011sq}, known as chromo-natural inflation~\cite{Adshead:2012kp, Adshead:2016omu, Bagherian:2022mau}, with a combination of both ideas put forth recently~\cite{Tishue:2022vwc}. It is also possible to use thermally-induced~\cite{1604.08838, 1805.07186} or scalar-induced~\cite{Creminelli:2023aly}  friction, rather than gauge-induced, to remove unwanted fast roll.

Here we construct a model of Higgs inflation where the Higgs is a pNG boson, based on Ref.~\cite{2210.10735}.\footnote{An inflationary setup where the Higgs is a pNG was considered in Ref.~\cite{1611.04932}, although there the Higgs does not play the role as the inflaton.} We use an  effective field theory (EFT) framework, in the spirit of chromo-natural inflation, to keep the Higgs slowly rolling down the steep pNG potential. This friction mechanism is the new ingredient which obviates the need for the flattening method of all previous Higgs inflation models. With the aim of demonstrating the key inflationary dynamics and putting such dynamics in the context of a realistic pNG Higgs model, we consider two Higgs inflation variants of increasing complexity. Specifically, we first consider a toy minimal pNG Higgs Inflation model where the Higgs boson interacts with the weak isospin gauge fields. Second, we investigate a more complete setup, based on the ``littlest Higgs" model~\cite{Arkani-Hamed:2002ikv}, and thus examine effects of additional 
 gauge fields (which are present in all pNG Higgs models). For specific parameters, such realistic models, and its generalizations, are known solutions to the Higgs hierarchy problem~\cite{Arkani-Hamed:2001nha, Schmaltz:2005ky, Perelstein:2005ka, Cheng:2004yc}. This model then also  allows us to engage with the possibility that both the Higgs and inflaton hierarchy problems are resolved in the same manner. We note that simpler pNG Higgs models have been ruled out experimentally~\cite{Chen:2006dy,Hewett:2002px}. In all cases, we find that  a sufficient number of $e$-folds can be achieved, albeit with large EFT couplings. 

This work is organized as follows. We present a minimal pNG Higgs and review the littlest Higgs model in Sec.~\ref{sec:full_setup}. In doing so, we demonstrate how our inflationary model can be implemented in an already-existing solution to the hierarchy problem. We also show how to obtain EFT couplings consistent with electroweak $SU(2)_{\rm L}\times U(1)_Y$ symmetry. Then, in Sec.~\ref{sec:higgsflation}, we present both the minimal pNG Higgs inflation model as well as its littlest Higgs variant and delineate the necessary model parameters to achieve successful inflation. We discuss potential methods to achieve large EFT couplings, lay out future directions of study, and conclude in Sec~\ref{sec:Discussion}.

{\sl Conventions and Notation}: We let $\hbar = c = 1$ and use $M_{\rm Pl}^2 = 1/(8\pi G)$ as the reduced Planck mass.  We let an overdot denote a derivative with respect to cosmic time, $\dot{f} = \tfrac{\intd f}{\intd t}$, and a prime denote a derivative with respect to the number of $e$-folds $N = \int \intd\log a$, $f' = \tfrac{\intd f}{\intd N}$, with $a$ the scale factor of the Friedmann-Lem\^aitre-Robertson-Walker (FLRW) metric $g_{\mu\nu} = a^2{\rm diag}(-1, 1, 1, 1)$. 
\section{\label{sec:full_setup}PNG Higgs}
We review the standard structure of the littlest-Higgs model and show how to express dimension-six Chern-Simons EFT couplings that obey the weak isospin $SU(2)_{\rm L}$ symmetry. Given the overall complexity of this model,  we begin in Sec.~\ref{sec:png_higgs} by displaying the minimal particle setup that will be required for inflation. We then address the entire particle-physics model in Sec.~\ref{sec:littlest}, delineating all new particles along with their interactions, calculating the form of the Higgs doublet potential, and concluding with the Chern-Simons EFT coupling. Readers interested in only the inflationary dynamics should go to Sec.~\ref{sec:higgsflation}.

\subsection{Minimal pNG Higgs}\label{sec:png_higgs}
We display a toy minimal pseudo-Nambu-Goldstone (pNG) Higgs model relevant for inflation that encompasses a wide variety of pNG Higgs models. The components common to this toy model consist of the usual Higgs doublet $H$, but now with a periodic potential $V(H)$ arising from higher-order dynamics. In addition, they contain a Higgs coupling to weak isospin gauge bosons through both covariant derivatives and a weak-isospin Chern-Simons current. Altogether, these terms take form in the Lagrangian
\begin{equation}
    \begin{split}
        \mathcal{L} &= D_{\mu}H^{\dagger}D^{\mu}H + V(H) + \frac{1}{4}W^{\mu\nu}_a W_{\mu\nu}^a + \frac{1}{4}B^{\mu\nu}B_{\mu\nu}
        \\
        & + \frac{g^2 \beta^2}{2f^2}\left(H^\dagger H\right)\text{Tr}\left[W^{\mu\nu}_a\tilde{W}_{\mu\nu}^a\right]
    \end{split}
    \label{eq:low_energy_Lagrangian}
\end{equation}
where $D_\mu = \partial_\mu - i g W_\mu^a \tau^a - (i/2)g' B_\mu$,  $g$ and $g'$ are the weak-isospin and hypercharge gauge couplings associated with bosons $W_\mu^a$ and $B_\mu$, respectively, and $\tau^a = \sigma^a/2$, with $\sigma^a$ the Pauli matrices. In addition, $W_{\mu\nu}^a = \partial_\mu W_\nu^a - \partial_\nu W_\mu^a + f_{ij}^a W_\mu^i W_\nu^j$ is the field-strength tensor associated with the weak-isospin bosons. Finally,  $\tilde{W}_{\mu\nu}^a = (1/2)\epsilon^{\mu\nu\alpha\beta}W_{\alpha\beta}^a$ is the weak-isospin field-strength tensor Hodge dual with $\epsilon^{\mu\nu\alpha\beta}$ the Levi-Civita tensor $[\tilde{\epsilon}^{\mu\nu\alpha\beta} = \sqrt{-{\rm det}\left(g_{\mu\nu}\right)}\epsilon^{\mu\nu\alpha\beta}$ is the flat-space Levi-Civita symbol$]$, and $f_{ajk} = \tilde{\epsilon}_{ajk}$ the $SU(2)$ structure constants. Generically-speaking, the Higgs potential $V(H)$ may be any periodic function; in the simplest case it is a cosine potential, but other calculable potentials are possible and depend on the specific model being considered, as will be seen in later sections of this work. We work in the unitary gauge, $H = (0, h/\sqrt{2})^\mathsf{T}$, with $h$ the background Higgs field. In this minimal setup, it will be the electroweak sector, by itself, that drives inflation.

\subsection{Littlest Higgs}\label{sec:littlest}
We focus on a model that can supply all of the components described in the previous section, known as the littlest Higgs~\cite{Arkani-Hamed:2002ikv, Perelstein:2003wd, Han:2003wu}. This model is based on a global $SU(5)/SO(5)$ coset and is the minimal enhancement to the Standard Model particle content to realize a viable pNG Higgs.  More specifically, the full theory we obtain is given by the Lagrangian
\begin{equation}\label{eq:full_L}
\begin{split}
\mathcal{L} &= \frac{f^2}{8}\mathrm{Tr}\left[\left(D_{\mu}\Sigma\right)^{\dagger}D^{\mu}\Sigma\right] - \frac{1}{4}\sum_{j = 1}^2 W^{\mu\nu}_{aj}W^{a j}_{\mu\nu}
\\
&- i\psi^{\dagger}\slashed{D}\psi - i\bar{u}^{\dagger}_i\slashed{D}\bar{u}^i + \text{H.c.}
\\
&-\frac{\lambda_1}{4}f\psi^{\dagger}_{i}\epsilon_{ijk}\epsilon_{xy}\Sigma_{jx}\Sigma_{ky}\bar{t}^{\dagger} + \lambda_2 f Ti\sigma_2 T + \text{H.c}
\\
&+\sum_{j=1}^2\frac{g_j^2 \beta^2}{16}\text{Tr}\left[Y\Sigma\left(Y\Sigma\right)^*\right]\text{Tr}\left[W^{\mu\nu}_{ja}\tilde{W}_{\mu\nu}^{ja}\right].
\end{split}
\end{equation}
To begin, recall that symmetries and symmetry breaking are key in many theories of particle physics. Little Higgs theories are generically built upon an interplay between spontaneously and explicitly broken global and gauged symmetries. This interplay mirrors the physics that describes pions as pseudo-Nambu-Goldstone bosons of a chiral flavor symmetry. A similar argument is invoked in little Higgs models, whereby the smallness of the Higgs's mass is due to some underlying strong dynamics, but the particular details of the underlying UV theory are negligible, instead encoding the physics in a semi-UV-complete theory. We first review its standard formulation in Sec.~\ref{subsec:slh} before detailing our additional components in Sec.~\ref{subsec:bslh}.

\subsubsection{Standard Littlest Higgs}\label{subsec:slh}
The littlest Higgs theory is characterized by a partial UV completion of the Standard Model using an approximate global $SU(5)$ symmetry along with a gauged $SU(2)\times[SU(2)\times U(1)]$ subgroup. We note that the original formulation of the littlest Higgs instead gauged a $[SU(2)\times U(1)]^2$ subgroup~\cite{Arkani-Hamed:2002ikv}. This subgroup differs from our choice by one factor of $U(1)$, i.e. it has an additional $U(1)$ gauge field. The removal of this additional $U(1)$ field does not affect the particular conclusions presented here~\cite{Perelstein:2003wd,Csaki:2003si}. In fact, the removal of the $U(1)$ gauge field relieves some of the phenomenological tensions encountered by the original littlest Higgs; we thus omit it. In terms of new field content, before global $SU(5)$ symmetry breaking, the littlest Higgs contains a collection of $11$ massless scalar fields, three massless gauge bosons, and a heavy vector-like fermion. We now describe how this new field content is realized.

 The 11 scalar fields, along with the 4 degrees of freedom in the Higgs doublet, are embedded within a symmetric $5\times 5$ matrix scalar $\Sigma$ under the {\bf 15} representation of $SU(5)$. Upon spontaneous symmetry breaking at an energy scale $\Lambda \sim 4\pi f$, this scalar obtains the vacuum expectation value (VEV) 
\begin{equation}
\Sigma_0 = \begin{pmatrix}	
& & \mathbb{1}_2
\\
 	&	1	&	
\\
\mathbb{1}_2	&		&	
\end{pmatrix},
\label{eq:vacuum}
\end{equation}
with $\mathbb{1}_N$ the $N\times N$ identity matrix and blank entries zeros. This VEV spontaneously breaks $SU(5)$ down to an $SO(5)$ subgroup. 

$SU(5)$ has 24 generators, while $SO(5)$ has 10 generators. It follows that $24 - 10 = 14$ massless scalars are produced in the particle spectrum, each corresponding to a broken generator of $SU(5)$. Moreover, the VEV in Eq.~\eqref{eq:vacuum} is chosen so that these 14 fields live in the fundamental representation of $SO(5)$ upon breaking. In addition to the 14 massless scalars, there is a single massive scalar that represents deviations from the VEV along the direction of symmetry breaking. We integrate this scalar out and make no further mention of it henceforth. The Lagrangian for the matrix scalar alone is that of a non-linear sigma model (NLSM):
\begin{equation}
\mathcal{L}_{\rm NLSM} = \frac{f^2}{8}\mathrm{Tr}\left[\partial_{\mu}\Sigma^{\dagger}\partial^{\mu}\Sigma\right],
\label{eq:Goldstone_Lagrangian}
\end{equation}
where 
\begin{equation}
\Sigma = \exp\left[\frac{2i}{f}\Pi\right]\Sigma_0 = \exp\left[\frac{2i}{f}T_a\pi_a\right]\Sigma_0
\label{eq:sigma_def},
\end{equation}
with $\Pi$ the `pion' field matrix and $T_a$ the unbroken $SU(5)$ generators ($f$ is also called `pion' decay constant). The pion matrix, fully written out, is
\begin{equation}
\Pi = \begin{pmatrix}
\omega - \frac{\eta}{\sqrt{20}}\mathbb{1}_2	&	\frac{{H}}{\sqrt{2}}	&	\phi^{\dagger}
\\
\frac{{H}^{\dagger}}{\sqrt{2}}	&	\sqrt{\frac{4}{5}}\eta	&	\frac{{H}^{\mathsf{T}}}{\sqrt{2}}
\\
\phi	&	\frac{{H}^*}{\sqrt{2}}	&	\omega^{\mathsf{T}} - \frac{\eta}{\sqrt{20}}\mathbb{1}_2
\end{pmatrix}.
\label{eq:Pion_matrix}
\end{equation}
As promised, we have a theory of 14 scalar fields: $\mathbf{\phi}$, a complex triplet, as a symmetric $2\times 2$ matrix, $H$ a complex doublet (our Higgs candidate), $\omega$, a real triplet, as a Hermitian $2\times2$ traceless matrix, and the real scalar $\eta$. The above fields transform with a shift symmetry under the broken generators of $SU(5)$, the specific field which does so depending on which broken generator is used for the transformation. Under the regime described thus far, the particles in $\Pi$ would be exactly massless at all scales, per the Goldstone theorem, with Lagrangian as in Eq.~\eqref{eq:Goldstone_Lagrangian}.

The symmetry is also \textsl{explicitly} broken, by gauging a $SU(2)\times [SU(2)\times U(1)]$ subgroup. 
This breaking is done by promoting the derivative in Eq.~\eqref{eq:Goldstone_Lagrangian} to a covariant derivative. In so doing, the Goldstone bosons become pseudo-Goldstone bosons that develop a mass proportional to the order parameter of the explicit breaking (which in this case are the gauge couplings of the gauged subgroup). This promotion also endows the Higgs with the requisite gauge couplings. Thus, Eq.~\eqref{eq:Goldstone_Lagrangian} becomes
\begin{equation}
\mathcal{L}^{\rm gauged}_{\rm NLSM} = \frac{f^2}{8}\mathrm{Tr}\left[\left(D_{\mu}\Sigma\right)^{\dagger}D^{\mu}\Sigma\right].
\label{eq:pseudo-NGB_Lagrangian}
\end{equation}

Here the covariant derivatives are of the form
\begin{equation}
D_{\mu}\Sigma = \partial_{\mu}\Sigma + ig_jW^{aj}_{\mu}\left(Q^j_a\Sigma + \Sigma Q_a^{j\mathsf{T}}\right) + ig'B_{\mu}\left(Y\Sigma + \Sigma Y\right)
\end{equation}
where $Q^j_a$ and $Y$ are the generators of the $SU(2)\times \left[SU(2)\times U(1)\right]$ gauged subgroup (i.e. $j \in \{1, 2\}$ and $a \in \{1, 2, 3\}$); these generators are written explicitly Appendix~\ref{app:EFT}.  With this promoted setup, the vacuum in Eq.~\eqref{eq:vacuum} spontaneously breaks the $SU(2)\times [SU(2)\times U(1)]$ symmetry down to the $SU(2)_{\rm L}\times U(1)_Y$ of the Standard Model prior to EWSB. When this breaking occurs, the SM couplings are recovered by the relation $g = \tfrac{g_1g_2}{\sqrt{g_1^2 + g_2^2}} = e/\sin \left(\theta_{\rm W}\right) = 0.64$ and $g' = e/\cos\left(\theta_{\rm W}\right) = 0.34$. Moreover, this breaking means that three of the generators are explicitly broken and thus become the longitudinal modes of the corresponding $W$ bosons post-symmetry-breaking. In other words, the $W_1$, $W_2$ bosons eat a combination of the three $\omega$ and $\eta$ NGBs to spit out three massive bosons (heavy counterparts of the SM $W$ bosons) and three massless bosons, which are the weak isospin bosons of the SM prior to EWSB. As a result, there is also a leftover massless Goldstone mode corresponding to the ungauged $U(1)$ subgroup; other works that have considered this model \cite{Csaki:2003si} have ignored the effects of this mode and we do so as well. The heavy gauge bosons have gauge couplings $g_H =  |g_1^2 - g_2^2|/\left(2\sqrt{g_1^2 + g_2^2}\right)$. The $U(1)$ gauge boson corresponds to our hyperweak $U(1)_Y$. The kinetic terms for the $SU(2)\times [SU(2)\times U(1)]$ gauge bosons are added in the canonical manner, which we label by $\mathcal{L}_{\rm kin}^{\rm gauge}$.

Coset models, such as this one, are also distinguished by their implementation of the coupling between the Higgs and top quark. The top is special because its large Yukawa coupling could potentially disrupt the naturalness of the Higgs potential. Other quarks, and leptons, are typically included according to their SM description. The littlest Higgs model seeks a minimal implementation (meaning the least number of new fermions). Hence, the $\Sigma$ field can be coupled to the top quark through a term of the form
\begin{equation}
\mathcal{L}_{\text{top}} = -\frac{\lambda_1}{4}f\psi^{\dagger}_{i}\epsilon_{ijk}\epsilon_{xy}\Sigma_{jx}\Sigma_{ky}\bar{t}^{\dagger} + \lambda_2 f Ti\sigma_2 T + \text{H.c},
\end{equation}
with $i, j, k \in \{1, 2, 3\}$ and $x, y \in  \{4, 5\}$. Here, $\psi$ is an $SU(3)$ triplet obtained by enhancing the usual left-handed third-generation quark doublet $q_3$ with a new heavy top, labelled $T$, 
\begin{equation*}
\psi_i = \begin{pmatrix}
-i\sigma_2 q_3
\\
T
\end{pmatrix} = \begin{pmatrix}
b
\\
t
\\
T
\end{pmatrix}.
\end{equation*}
Note the presence of two couplings, $\lambda_1$ and $\lambda_2$, as well as the Dirac mass term for the new heavy top $T$. The top quark mass eigenstate will be a mixture of the new heavy top and the usual left-handed quark doublet top. As a result, the Higgs couples neither to $\lambda_1$ nor $\lambda_2$, but to a product $\lambda_1\lambda_2$.
\begin{equation}
\mathcal{L}\supset \lambda_tq_3h\bar{t}\qquad\text{where}\qquad\lambda_t = \frac{\lambda_1\lambda_2}{\sqrt{\lambda_1^2 + \lambda_2^2}}
\end{equation}
Thus if either coupling is turned off, the Higgs receives no quadratic divergence from the top quark at all. This lack of divergence is a manifestation of the generic collective symmetry breaking pattern of little Higgs theories. When either coupling is turned off, the theory has an enhanced global  $SU(3)$ symmetry, which renders the coupling technically natural.

The other fermions of the Standard Model may be incorporated in the usual fashion; although this does violate the collective symmetry breaking structure that prevents quadratic divergences in the scalar sector, the relative lightness of all other fermions compared to the Higgs that such divergences are acceptable and to some extent negligible.

Now we must find the potential that the Higgs field experiences. Note that most treatments of the little Higgs only consider the effective potential to leading or next-to-leading order in the Higgs field $h$. Here we instead consider a full treatment, in order to illustrate the periodic potential experienced by the Higgs field, which plays an important role in the inflationary dynamics.

A one-loop analysis yields additional operators that must be added to the Lagrangian in Eq.~\eqref{eq:pseudo-NGB_Lagrangian} in order to account for quadratic divergences. These terms are of the form
\begin{equation}
\begin{split}
    \mathcal{L} &\supset cf^2\sum_{j = 1}^2 g_j^2{\rm Tr}\left[\left(Q^j_a\Sigma\right)\left(Q^j_a\Sigma\right)^*\right]
    \\
    &+ cf^2g' {}^2\text{Tr}\left[\left(Y\Sigma\right)\left(Y\Sigma\right)^*\right]
    \end{split}
\label{eq:gauge_Weinberg_terms}
\end{equation}
from gauge boson loops and
\begin{align}
\mathcal{L} \supset 
 c'\lambda_1\epsilon_{wx}\epsilon_{yz}\epsilon_{ijk}\epsilon_{\ell mn}\Sigma_{iw}\Sigma_{jx}\Sigma^*_{my}\Sigma^*_{nz} + {\rm H.c.},
\label{eq:fermion_Weinberg_terms}
\end{align}
from fermion loops. Here, $c$ and $c'$ are undetermined coefficients, consistent with Wilson renormalization analysis, whose exact value depends on the UV completion. In the unitary gauge, the VEV of the $\Sigma$ field $\Braket{\Sigma}$ may be written as
\begin{equation}
    \Braket{\Pi} = \begin{pmatrix}
      &   \Braket{H}  &  
    \\
    \Braket{H}^{\dagger}   &     &   \Braket{H}^{\mathsf{T}}
    \\
      &   \Braket{H}^*  &   
    \end{pmatrix}.
    \label{eq:pi_vac}
\end{equation}
where the Higgs doublet $H = \left(0, h/\sqrt{2}\right)^\mathsf{T}$ (n.b. defined as a column vector). 
Using Eq.~\eqref{eq:sigma_def}, we can write the VEV of $\Sigma$ as a function of the Higgs boson $h$. Using the expression for $\Braket{\Pi}$ from Eq.~\eqref{eq:pi_vac} allows us to write
\begin{equation} 
    \begin{split}
        \Braket{\Sigma} = \left[\mathbb{1}_5 + \frac{i}{h}\sin\left(\frac{h}{f}\right)\Braket{\Pi} - \frac{2}{h^2}\sin^2\left(\frac{h}{f}\right)\Braket{\Pi}^2\right]\Sigma_0
    \end{split}
    \label{eq:Sigma_VEV}
\end{equation}
Plugging $\Braket{\Sigma}$ into the terms shown in Eq.(\ref{eq:gauge_Weinberg_terms}-\ref{eq:fermion_Weinberg_terms}) yields a tree-level potential for $h$ \cite{Perelstein:2003wd,Espinosa:2004pn}:
\begin{equation}\label{eq:higgs_potential}
    V_0(h) = D + f^4\left[cg'^2\sin^2\left(\frac{2h}{f}\right)+\frac{1}{2}\lambda_+\sin^4\left(\frac{h}{f}\right)\right]
\end{equation}
where $D/f^4$ is an $\mathcal{O}\left(10^{-2}\right)$ constant and $\lambda_{\pm} = c\left(g_1^2\pm g_2^2\right)\pm16c'\lambda_1^2$. This potential is periodic in $h$, and electroweak symmetry breaking (EWSB) does not take place, as can be seen by expanding the potential about $h\rightarrow 0$; this feature is in keeping with the expectation of the Higgs as a Goldstone boson. A small non-zero Higgs mass proportional to the $U(1)_Y$ coupling $g'$ does arise at this level, due to the explicit breaking of $SU(5)$ by the gauging of the $U(1)_Y$ subgroup. It can be shown that this mass must be positive, meaning EWSB does not take place at tree level \cite{Espinosa:2004pn}.

EWSB instead arises through the Coleman-Weinberg mechanism, resulting in a negative loop-level Higgs mass and self-coupling terms. The full VEV of Eq.~\eqref{eq:Sigma_VEV} can be used to compute the mass matrices for the gauge bosons $M_W$, fermions $M_t$ and remaining scalar fields $M_s$, in particular the triplet field $\phi$ \cite{Perelstein:2003wd}. Note that these expressions will be periodic in $h$, in accordance with its approximate shift symmetry. These matrices can be used to compute the CW contributions which remain once the heavy fields (the top partner, the Higgs triplet, etc) have been integrated out, which have the form \cite{Kilic:2015joa}:
\begin{align}
V_{\rm CW} = \frac{1}{64\pi^2}\text{STr}\left[\left(M^{\dagger}M\right)^2\left(\log\left(\frac{M^{\dagger}M}{\Lambda^2}\right)-\frac{3}{2}\right)\right]
\end{align}
The supertrace takes into account both the statistics [$(+1)$ for bosons, $(-1)$ for fermions] and the multiplicity of the fields (e.g., the weak boson contributions are multiplied by a factor of three for the three polarizations). The inclusion of these contributions leads to a negative Higgs mass term and consequently to EWSB. 
\begin{figure}
    
    \includegraphics[width=0.45\textwidth]{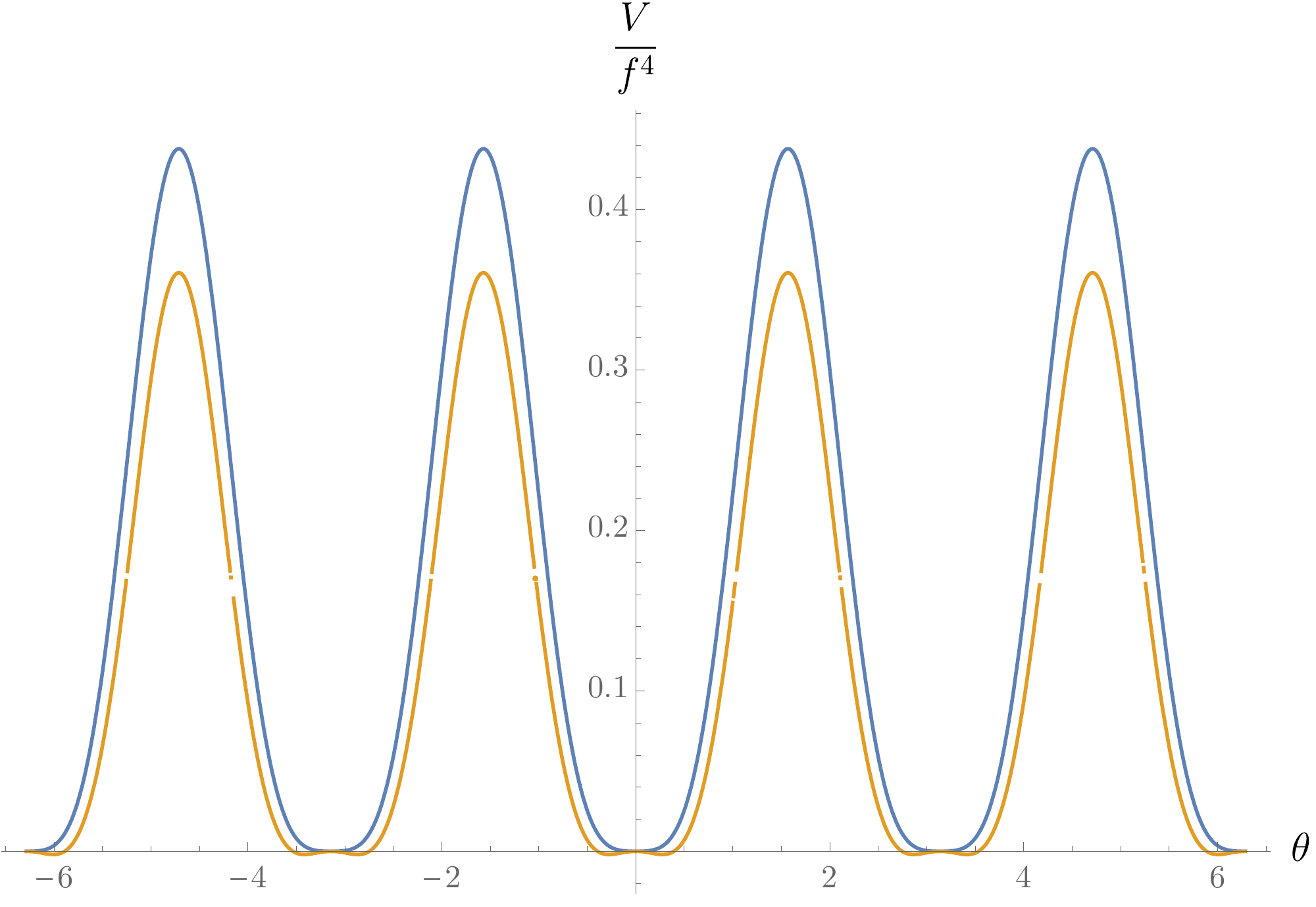}
    \caption*{(a)\label{fig:CW_potential}}
    
\hfill
    
    \includegraphics[width=0.45\textwidth]{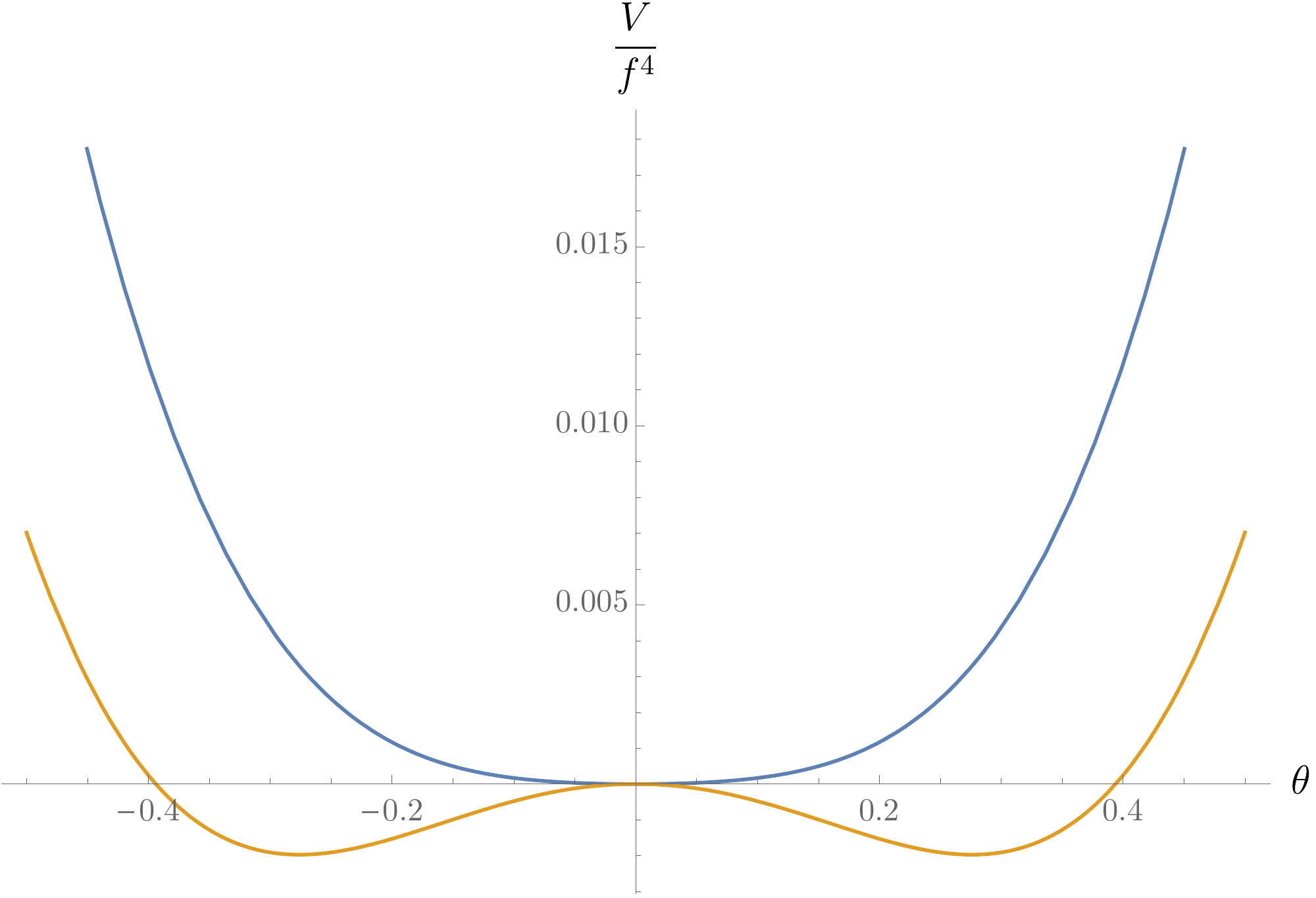}
    \caption*{(b)\label{fig:CW_potential-expansion}}

    \caption{\label{fig:potentials} The periodic potential of the pNGB Higgs in the $SU(5)/SO(5)$ coset scenario as a function of $\theta = \frac{h}{f}$. The blue lines correspond to the periodic potential generated by the Higgs VEV; the orange line includes the contributions from the Coleman-Weinberg terms. The full potential can be seen in the top figure. The potential near the origin, exhibiting EWSB, is shown in the bottom figure. Both figures are made using fiducial choices of parameters $g' = 0.34$, $g_1 = g_2 = \sqrt{2}g = 0.91$, $c = 0.05$, $c' = -0.05$ and $\lambda_1 = 1$ (so that $\lambda_+ = -0.312$). }
\end{figure}
The contributions are from the heavy top quark, the heavy $W$ boson, and the integrated-out Higgs triplet, respectively. These two potentials are shown in Fig.~\ref{fig:potentials}.

\subsubsection{Beyond Standard Littlest Higgs}\label{subsec:bslh}
The Higgs has an observed mass of $m_H = 125\ {\rm GeV}$~\cite{ParticleDataGroup:2022pth}. Hence, for $f\gg 10\ {\rm TeV}$, the tree-level potential in Eq.~\eqref{eq:higgs_potential} does not predict the correct Higgs mass. In order to explore a larger phenomenological range of decay constants, while adhering to observations, we make a small change, $f^4 \rightarrow \mu^4$, leading to the following potential
\begin{align}\label{eq:BSLHP}
V_0(h) = \mu^4\left[cg'^2\sin^2\left(\frac{2h}{f}\right)+\frac{1}{2}\lambda_+\sin^4\left(\frac{h}{f}\right)\right],
\end{align}
with $\mu$ a new parameter that satisfies 
\begin{equation}
m_H^2 = 8 c g'^2 \left(\mu^2/f\right)^2
\end{equation}
so that $\mu/f \ll 1$ for the regime of interest (e.g. in Sec.~\ref{subsec:min_inf} we will show that we require $f\gtrsim 10^9$ GeV implying $\mu/f \lesssim 10^{-3}$).  Such a scale difference between the width and height of a potential, required by the amplitude of density perturbations arising in inflation, is common in axionic theories, where $\mu$ would roughly be the strong coupling scale and $f$ the axion decay constant~\cite{Marsh:2015xka, Hook:2019qoh}. Hence, for convenience, we refer to $\mu$ as the strong coupling scale. In our case, we posit that such a difference could occur once a full UV completion of the Higgs is considered (i.e. in some composite Higgs model that embeds the little Higgs structure).

In addition to the change in potential, we also employ an EFT analysis to include two dimension-six Chern-Simons terms coupled to the scalar fields. This term has the form
\begin{equation}\label{eq:Chern-Simons}
\mathcal{L}_{\rm CS}  = \sum_{j = 1}^2 \frac{g_j^2\beta^2}{16}\text{Tr}\left[Y\Sigma\left(Y\Sigma\right)^*\right]\text{Tr}\left[W^{\mu\nu}_{ja}\tilde{W}_{\mu\nu}^{ja}\right],
\end{equation}
where we take both gauge fields to have the same dimensionless Chern-Simons coupling $\beta$ and we explicitly point out that $*$ denotes complex (not Hermitian) conjugation. The invariance of this expression under gauge transformations is explicitly shown in Appendix~\ref{app:EFT}. Upon inserting the expansion of the $\Sigma$ field, the lowest-order term gives the usual Chern-Simons factor, which is a total derivative. The higher-order terms yield couplings between the Chern-Simons current and the scalar sector of the theory. 
\section{pNG Higgs Inflation}\label{sec:higgsflation}
Given the particle Lagrangians in Sec.~\ref{sec:full_setup}, we now turn to their inflationary dynamics, i.e. we both solve for the dynamics and quantify the necessary Lagrangian parameters to obtain successful pNG Higgs inflation. By successful pNG Higgs inflation, we mean that at least $N \sim 60$ $e$-folds can be achieved. To do so, we again first consider the minimal inflationary setup displayed in Sec.~\ref{subsec:min_inf} and then its littlest-Higgs variant in Sec.~\ref{subsec:little_inf}.
\subsection{Minimal pNG Higgs Inflation}\label{subsec:min_inf}

We begin with the Lagrangian in Eq.~\eqref{eq:low_energy_Lagrangian} and, for simplicity, take a Higgs potential of the form
\begin{equation}
V(\theta) = \mu^4\left[1 + \cos\left(\theta\right)\right],
\end{equation}
where $\mu$ is the amplitude of the potential and $\theta = \frac{h}{f}$ is the normalized Higgs boson. In this case, the Higgs mass is $m_H = (\mu^2/f)^2$. If we treat $f$ as the scale of spontaneous symmetry breaking, then we take $f\ll M_{\rm Pl} $ in order to safely neglect quantum gravity corrections to our Lagrangian.\footnote{If instead we take $f/\beta$ as the scale of spontaneous symmetry breaking, then it is possible to have $f\gtrsim M_{\rm Pl}$. In this case, $f \ll M_{\rm Pl}$ is an assumption in the following treatment. 
} In this limit, the Higgs-covariant derivative $D_\mu$ can be treated as a flat-space derivative $\partial_\mu$ since the Higgs interactions with weak isospin gauge fields are suppressed by a factor $\left(f/M_{\rm Pl}\right)^2\ll 1$. 

In order to maintain isotropy of the Universe, the inflaton is only a function of time, $h(t, {\bf x}) = h(t)$. For the same reason, a classical and rotationally-invariant attractor gauge-field configuration is chosen as in chromo-natural inflation~\cite{Adshead:2012kp}: 
\begin{align}
W_0^a(t) = 0&,\quad W_i^a(t) = a(t)\psi(t)\delta_i^a,\\
W_{0i}^a(t) = \partial_t[a(t)\psi(t)]\delta_i^a&,\quad W_{ij}^a(t) = -g f_{ij}^a[a(t)\psi(t)]^2.
\end{align}
A general gauge-field configuration redshifts away its anistropic parts during inflation, as the Chern-Simons term is only sensitive to the isotropic piece, so that the rotationally-invariant configuration can be dynamically achieved~\cite{1311.3361, 2003.01617, 2105.06259}. The hypercharge gauge boson is assumed to be identically zero, $B_\mu = 0$.

Successful inflation occurs when the Hubble parameter $H$ slowly evolves, $\varepsilon = -\dot{H}/H^2 \ll 1$. Under this slow-roll condition, the Friedmann equations are
\begin{align}
 H^2 &= \frac{\mu^4}{3M_{\rm Pl}^2}\left[1 + \cos(\theta)\right],\\
\varepsilon &=  \varepsilon_h + \varepsilon_\psi\left(1 + 2\eta_\psi + \eta_\psi^2 + m_\psi^2\right),
\end{align}
where we have introduced the dimensionless mass parameter  $m_\psi =\frac{g\psi}{H}$ and where
\begin{align}
\varepsilon_h &= \frac{\dot{h}^2}{2M_{\rm Pl}^2H^2},\quad \varepsilon_\psi = \left(\frac{\psi}{M_{\rm Pl}}\right)^2 ,\quad \eta_\psi = \frac{\dot{\psi}}{H\psi},
\end{align}
are slow-roll parameters~\cite{Adshead:2016omu}. Since inflation requires $\varepsilon < 1$,  each term in $\varepsilon$ must also be small. In addition, for the slow-roll solution to persist, the acceleration of the fields must also be small,
\begin{align}
-\frac{\ddot{h}}{H\dot{h}} = \varepsilon +\hat{\delta}_\psi \ll 1&,\quad  -\frac{\ddot{\psi}}{H\dot{\psi}} = \varepsilon +\hat{\delta}_\psi \ll 1,\\
\implies \hat{\delta}_h = -\frac{h''}{h'} \ll 1&, \quad \hat{\delta}_\psi = -\frac{\psi''}{\psi'} \ll 1,
\end{align} 
where overdot indicates a cosmic time derivative and prime a conformal time derivative. With the above conditions, the slow-roll equations of motion for both the Higgs and the gauge-fields are then
\begin{align}\label{eq:slow_th}
3\theta' - A\sin(\theta) = -3g^2\frac{\beta^2}{\left(f/M_{\rm Pl}\right)^2}  \varepsilon_\psi m_\psi \theta,\\
3\varepsilon_\psi' + 4(1 + m_\psi^2)\varepsilon_\psi = 2g^2\beta^2 m_\psi \theta \theta' \varepsilon_\psi,
\end{align}
with $A = \mu^4/(H^2 f^2)$ and we have traded the gauge field $\psi$ for its slow-roll parameter $\varepsilon_\psi$. The left-hand sides of these equations are equivalent to those in Ref.~\cite{Adshead:2012kp}, rewritten in terms of $\varepsilon_\psi$. The right-hand sides are similar, but different, due to a dimension six (rather than a dimension five) Chern-Simons operator in Eq.~\eqref{eq:Chern-Simons} with dimensionless Chern-Simons coupling $\beta$. We seek static gauge-field solutions, $\varepsilon_\psi' = 0$, so that the second of the above equations becomes
\begin{align}
\theta' &= \frac{2}{g^2\beta^2}\frac{1 + m_\psi^2}{\theta m_\psi}.
\end{align}
Combining this equation with Eq.~\eqref{eq:slow_th} yields
\begin{align}
\varepsilon_\psi m_\psi \theta &= \frac{\left(f/M_{\rm Pl}\right)^2}{\beta^2}\left[\frac{1}{3}A\sin(\theta) - \frac{2}{g^2\beta^2}\frac{1 + m_\psi^2}{ m_\psi \theta}\right].
\end{align}
This equation has a simple solution by further assuming $g^2\beta^2 m_\psi A \theta \sin(\theta)\gg 1$. That is, noting that $H^2 m_\psi^2 = g^2 M_{\rm Pl}^2 \varepsilon_\psi$, we get
\begin{align}
\varepsilon_\psi &= \frac{\rho^{1/3}}{g^2 \beta^2}\left\{\frac{{\rm sinc}(\theta)}{[1 + \cos(\theta)]^{1/2}}\right\}^{2/3},
\end{align}
with $\rho = \left(\beta^2/3\right)\left(\mu/M_{\rm Pl}\right)^4$ and ${\rm sinc}(\theta) = \sin(\theta)/\theta$. If inflation begins at the top of the potential and ends at the bottom, so that $\theta \in [0, \pi]$, then it lasts for 
\begin{align}
\nonumber N &= \int_{0}^{\pi} \frac{d\theta}{\theta'} = \frac{g^2\beta^2}{2}\int_0^\pi d\theta\ \theta\left[\frac{m_\psi(\theta)}{1 + m_\psi^2(\theta)}\right]\\
&=  \frac{g^2\beta^2}{2}\int_{0}^{\pi} d\theta\ \theta\frac{\rho^{1/3}[1 + \cos(\theta)]^{2/3}{\rm sinc}^{1/3}(\theta)}{\rho^{2/3}[1 + \cos(\theta)]^{4/3} + {\rm sinc}^{2/3}(\theta)},
\end{align}
$e$-folds. The maximum number of $e$-folds occurs around $\rho \approx 1$ and, in this case,  $N\approx g^2\beta^2$. If instead $\rho \ll 1$, which is the regime of interest, then
\begin{align}\label{eq:N}
N &\approx 60\ \left(\frac{g}{0.64}\right)^2\left(\frac{\beta}{3\times 10^6}\right)^{8/3}\left(\frac{\mu}{8 \times 10^6\ {\rm GeV}}\right)^{4/3}.
\end{align}

\begin{figure}
\includegraphics[width=0.4\textwidth]{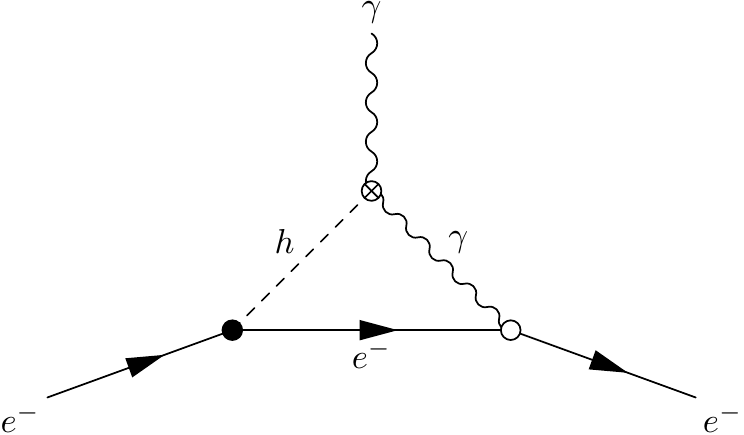}
 \caption{Triangle diagram for the modified electric dipole moment (EDM) of the electron in our model. The solid arrow lines represent electrons, the wavy lines photons, and the dashed line the Higgs boson (post-EWSB). The black dot is the Yukawa electron-Higgs interaction, the empty dot the covariant-derivative electron-photon interaction, and the crossed dot the induced electromagnetic Higgs-Chern-Simons interaction described in Appendix~\ref{app:eEDM}. }\label{fig:eEDM}
 \end{figure}

In addition to the above $e$-fold constraint, there is an additional constraint from the electric dipole moment (EDM) of the electron $d_e$~\cite{Lue:1996pr, Zhang:1993vh}. More precisely, the dimension-six Chern-Simons operator induces this EDM through the triangle diagram in 
 Fig.~\ref{fig:eEDM}.
Current bounds on this EDM  come from spin precession measurements of polar thorium monoxide (ThO) molecules by the Advanced Cold Molecule Electron EDM (ACME) collaboration, which provide the limit $d_e \lesssim 1.1\times 10^{-29} e\ {\rm cm}$~\cite{ACME:2018yjb}.

\begin{figure}[ht]
\includegraphics[width=0.45\textwidth]{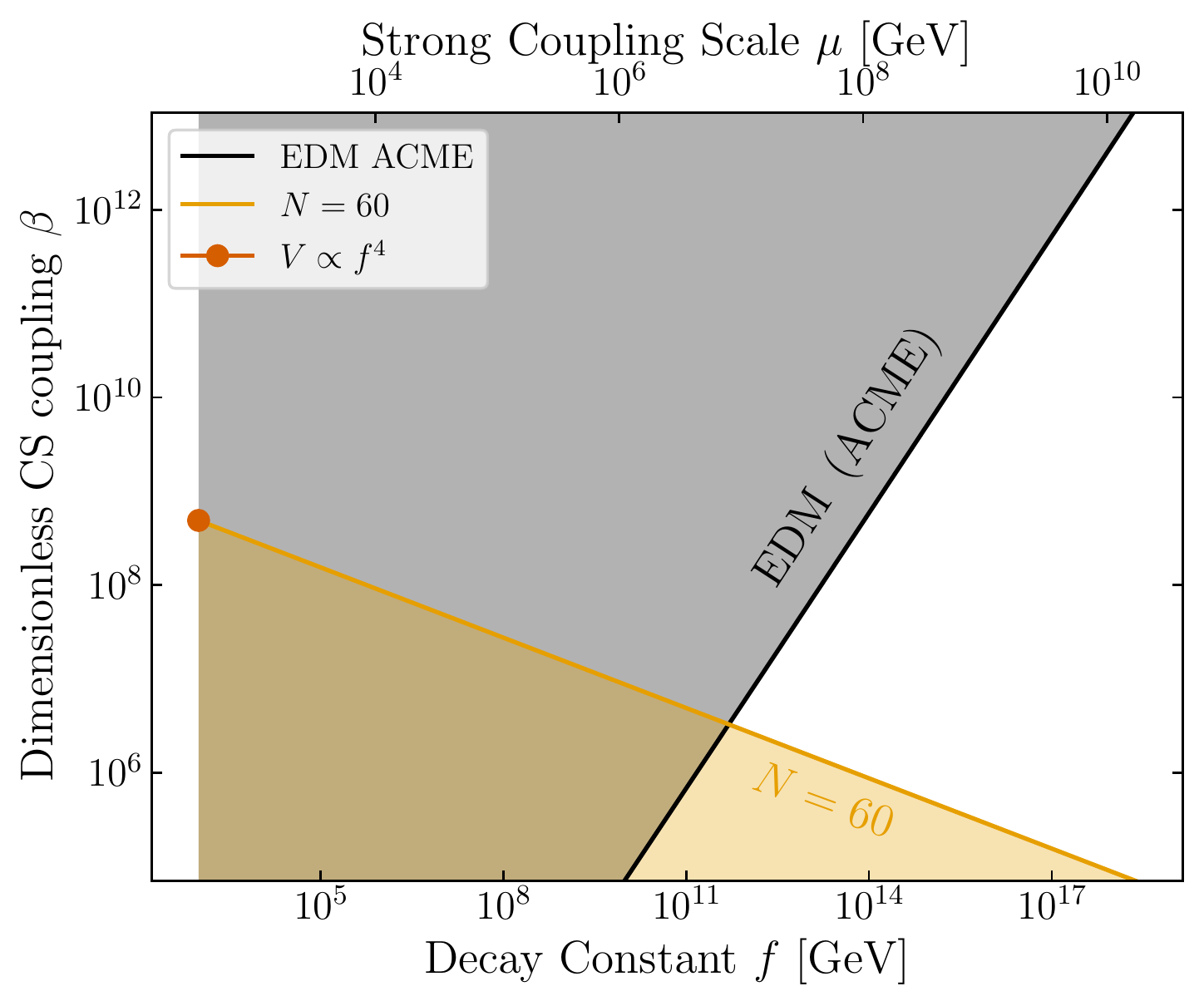}
\caption{The viable parameter space for minimal pNG Higgs Inflation, assuming sub-Planckian decay constants, is shown in white. The lower horizontal axis is the $SU(5)$ decay constant $f$, the corresponding upper horizontal axis is the potential height $\mu$ [given by the relation 
$m_H^2 = \left(\mu^2/f\right)^2$], and the vertical axis is the dimensionless Chern-Simons (CS) coupling $\beta$. The black line is the maximum couplings allowed by electric-dipole moment (EDM) constraints. The orange line is the minimum required couplings to achieve $N = 60$ $e$-folds of inflation.  These two constraints imply that only decay constants satisfying $f \gtrsim 5\times 10^{11}\ {\rm GeV}$ are not ruled out. The red dot indicates the predicted regime given by the standard littlest Higgs potential with $V \propto f^4$ instead of $V \propto \mu^4$. 
}\label{fig:param_space}
\end{figure}

We show both of these constraints in Fig.~\ref{fig:param_space}. The decay constant shown in the lower horizontal axis must satisfy $f \gtrsim 5\times 10^{11}\ {\rm GeV}$; the relation $m_H^2 = \left(\mu^2/f\right)^2$ then leads to the values of $\mu$ as shown in the upper horizontal axis. The dimensionless Chern-Simons coupling shown in the vertical axis must satisfy $10^5 \lesssim \beta \lesssim 10^{13}$. 

\subsection{Littlest Higgs Inflation}\label{subsec:little_inf}
There are two main differences between the minimal pNG Higgs and littlest Higgs. First, the Higgs potential is different. Second, there are additional fields. 

If the Higgs potential takes the form $V(\theta) = c_0 + c_2\sin^2\left(2 \theta\right) + c_4\sin^4\left(\theta\right)$, as in the littlest Higgs model, we numerically find that the number of $e$-folds is roughly given by Eq.~\eqref{eq:N}, up to small order unity corrections. We also point out that $8c g'^2 \approx 1$, with $c = 1$, so that the minimal PNG and littlest-Higgs models have the same parameterization of the Higgs mass, $m_H \approx (\mu^2/f)^2$.

The main additional fields during the inflating phase (i.e. after global $SU(5)$ symmetry breaking) are the heavy $SU(2)$ bosons. We now address whether the heavy gauge boson can be dynamical. A massive gauge boson with mass $M_H$ changes all the equations of motions from $m_\psi^2 \rightarrow m_\psi^2 + M_H^2/H^2$. For slowly-rolling gauge fields, the fast oscillations of the mass term dominate the energy density and lead to a suppressed number of $e$-folds by a factor of the squared heavy-boson mass $M_H^2/H^2 \sim \left(f/M_{\rm Pl}\right)^2$. As a result, $\beta$ must be even larger to compensate for the smallness of the boson mass. To avoid such a situation, we therefore conclude that the heavy gauge-boson must have zero dynamics, so that littlest-Higgs inflation reduces to the minimal set up in Sec.~\ref{subsec:min_inf}.

\section{\label{sec:Discussion}Discussion and Conclusion}
Why should the Higgs be the inflaton? One line of reasoning is that its presence as the only scalar in the SM yields a minimal explanation for the identity of the inflaton. Another is that the inflaton and Higgs boson both suffer from a hierarchy problem: loop corrections to their respective potentials spoil the viability of their theories. Previous work on Higgs inflation has led to a rich series of models, rooted in modifications of gravity, whose predictiveness is highly dependent on the precise resolution to their violations of perturbative unitarity, which is largely unknown. Here, we thus engaged with an alternative realization for Higgs inflation without modifying gravity, whereby the Higgs has an approximate shift-symmetry as a pNG boson and slow-roll is sustained through friction induced by a suitable weak-isospin Chern-Simons operator. In our scenario, inflation thus happens only within the electroweak sector.

In order to maintain a general framework, we consider two variants of pNG Higgs inflation. One, where we use only the minimal ingredients as dictated by our friction-assisted pNG description and another that embeds this minimal model by explicitly placing the pNG Higgs into a non-linear sigma model, i.e. we consider a little Higgs embedding. Specifically, we choose the simplest of such models, the littlest Higgs.  Moreover, in order to consider the widest possible range of inflationary scales, we make a phenomenological replacement and consider the height and width of the Higgs potential to be independent parameters. However, we note that the standard description for the littlest Higgs only works around $f\sim$TeV scales; larger values of $f$ wreck the cancellations of quadratic divergences and restore the hierarchy problem. Nevertheless, we imagine that our amplitude replacement is reasonable, as similar scalings are seen in other pNG potentials (e.g. axions), and surmise that it can be obtained in some true UV completion of the littlest Higgs which would fully address the hierarchy problem (e.g. some composite Higgs theory). We leave investigation into such UV completions for future work.

We find that, in both cases, successful Higgs inflation can occur at high energies, with decay constants $f \gtrsim 5\times 10^{11}\ {\rm GeV}$, and with a large dimensionless Chern-Simons couplings $10^5 \lesssim \beta 10^{13}$ (see Fig.~\ref{fig:param_space}). It may be possible to achieve this large coupling through a variety of means, such as by including other couplings between the Higgs and the gauge fields, as in models with deconstructed dimensions~\cite{Arkani-Hamed:2001kyx}, or including multiple Higgs fields, such as in assisted~\cite{Liddle:1998jc}, $N$-flation~\cite{Dimopoulos:2005ac}, or clockwork~\cite{Kaplan:2015fuy, Craig:2017cda, Agrawal:2018mkd} models. The additional fields (e.g. the massless $\eta$) may serve as one such assisted field when properly considered. In the case of multiple Higgs, we would be considering multiple copies of the model at hand.  It may also be possible to lessen the large dimensionless coupling with models that have additional sources of friction, such as in warm inflation models ~\cite{Berera:1995wh,Berera:1995ie, Visinelli:2011jy, Berghaus:2019whh, Montefalcone:2022jfw}.  

Instead of using SM field content, a dark Higgs and/or dark $SU(2)$ gauge-fields can instead be used to achieve inflation. In the case of a dark Higgs, the calculations we have presented remain the same, although the restriction of recovering the SM Higgs mass and TeV electroweak physics is lifted. We note that, in either case, it also is possible to use the little Higgs framework in concert with the original non-minimal coupling of Higgs inflation, rather than a new  source of friction. 

We have primarily focused here on the background evolution of inflation and largely ignored predictions of the perturbations. Given the similarities between this model and that of chromo-natural models, one immediate concern is that 
those models are not in agreement with observations of the cosmic microwave background; the original chromo-natural construction fails to produce the correct spectral index for the primordial scalar power spectrum and does not satisfy experimental constraints on the tensor-to-scalar ratio. However, the original authors found that introducing a mass term for the gauge field (by Higgsing) screens the gauge field fluctuations thus rendering chromo-natural inflation observationally viable~\cite{Adshead:2016omu}.  In our model, since the Higgs obtains a VEV during inflation, it naturally induces a mass for the gauge fields and is likewise viable, by extension.  We will consider a full cosmological perturbation analysis for future work.  

Moreover, while the additional fields of littlest Higgs inflation do not change the dynamics of the background evolution, they can have an impact on the perturbations through non-Gaussian signatures~\cite{Lee:2016vti}. That is, the additional fields in question have masses near the scale of inflation, leading to oscillatory features in cosmological correlators generic to quasi-single field inflation~\cite{Chen:2009zp}. In addition, interactions between the Higgs and these other fields will yield cosmological collider signatures~\cite{Arkani-Hamed:2015bza,  Lu:2019tjj}.

This work lays the foundation for further studies with this model and its extensions. For example, in one future work we will examine this model through the lens of the analysis done in Ref.~\cite{DeRocco:2021rzv}; there exists the possibility that this model, like many other models that employ a natural inflation-like mechanism, must inherently take backreaction effects into account, thus making them models of warm inflation. As mentioned previously, this warming of inflation may be an avenue to lower the dimensionless Chern-Simons coupling, making it a logical extension.

A simplifying assumption that underpins this work is that the Higgs rolls into the same EW-breaking minimum throughout the entire universe; a priori this need not be the case. Given that Higgs-dependent masses are $2\pi$-periodic in $H$, these various vacua would be indistinguishable experimentally, but would have lasting signals such as domain walls or tunnelling phenomena.  

Finally, our approach has the virtue of providing a natural baryogenesis~\cite{Lue:1996pr} and reheating mechanism, since the Higgs inflaton has gauge invariant couplings to leptons and baryons and will decay into baryons, leptons and gauge bosons at the end of inflation.  A detailed analysis of such a baryogenesis and reheating mechanism inherent to our model will also be pursued in the near future.
\subsection*{Acknowledgments}
We thank Aaron Pierce and Robert McGehee for useful discussions. KF thanks Can Kilic and Barmak Shams for extremely helpful conversations. CCS thanks Ken Van Tilburg for reading a previous draft and providing useful comments.  SA was supported by the Simons Foundation award number 896696. HBG is supported by the National Science Foundation, MPS Ascending Fellowship, Award 2213126. KF is grateful for support from
the Jeff and Gail Kodosky Endowed Chair in Physics at
the Univ. of Texas, Austin. KF acknowledges funding
from the U.S. Department of Energy, Office of Science,
Office of High Energy Physics program under Award
Number DE-SC0022021. KF acknowledges support by
the Vetenskapsradet (Swedish Research Council) through
contract No. 638- 2013-8993 and the Oskar Klein Centre
for Cosmoparticle Physics at Stockholm University.
\appendix
\section{Pion Physics}
In this appendix we briefly review the physics of pions, as a reminder of the structures employed in the construction of little Higgs models. At low energies, the fundamental degrees of freedom of QCD are confined and reorganized into composite degrees of freedom, namely the baryons and mesons.

At low energies ($\sim100$ MeV), of particular relevance are the up, the down, and, to a lesser extent, the strange quarks. Compared to their charm, bottom, and top siblings, these quarks are effectively massless ($m_u\sim 2$ MeV, $m_d\sim 4$ MeV, $m_s\sim95$ MeV). This (relative) degeneracy means that the effective theory has a flavor symmetry, $SU(2)_V$ [$SU(3)_V$ if the strange quarks are included in the discussion; we'll proceed with just the up/down case for this section].

Suppose that the quarks are taken to be massless. In this case, the symmetry is exact and is furthermore enlarged to a chiral symmetry $SU(2)_{\text{L}}\times SU(2)_{\text{R}}$. This symmetry has $2\times (2^2-1) = 6$ generators. Of course, quarks are not massless; dimensional transmutation in QCD introduces a mass scale ($\Lambda_{\rm QCD}$), so the massless limit chiral symmetry is spontaneously broken into the diagonal flavor symmetry as above, $SU(2)_V$, which has $3$ generators. Thus, by the Goldstone theorem, we would expect there to be 3 massless scalar modes in the spectrum of the theory. Furthermore, these Goldstone modes remain massless to all orders, which limits them to derivative couplings that may be written down in the Lagrangian. In this case, these three modes are known as the pions, $\pi^{\pm}, \pi^0$. The dynamics of the pions are encoded as a non-linear sigma model (NLSM), in a Lagrangian of the form
\begin{equation}
    \mathcal{L} = \frac{f^2}{8}\text{Tr}\left[\partial^{\mu}U^{\dagger}\partial_{\mu}U\right]
\end{equation}
where $U$ is a unitary operator $U = \exp\left[\frac{2\pi i}{f}\vec{\pi}\cdot\vec{T}\right]$. Here $\vec{T}$ is the vector of broken $SU(2)$ generators. Note that here it is very easy to see that under a transformation $\vec{\pi}\rightarrow\vec{\pi} + f\vec{\alpha}$ the Lagrangian is manifestly invariant.

Of course, the flavor symmetry is not exact; in addition to the spontaneous symmetry breaking, the symmetry is also broken explicitly by the Yukawa couplings which introduce mass differences between the quarks. The Goldstone theorem may be extended to such a case, wherein the Goldstone modes are no longer massless, but instead slightly massive, with the mass scale proportional to the degree to which the symmetry is explicitly broken. In this case, this means that the mass of the pions is controlled by the mass difference between the up and down quark. 

In little Higgs models, this structure is adapted; the overarching flavor symmetry is replaced by a larger symmetry group of some UV-complete theory. In the littlest Higgs this is the $SU(5)$ group. The breaking of this symmetry to some `intermediate' group, which happens spontaneously, corresponds in the pion framework to the $SU(2)_V$ diagonal symmetry and to $SO(5)$ in the littlest Higgs framework. Lastly, the explicit breaking of $SU(5)$ to the $SU(2)\times [SU(2)\times U(1)]$ subgroup is analogous to the breaking of $SU(2)_{\mathrm{L}}\times SU(2)_{\mathrm{R}}$ by the quark mass differences.

\section{Chern-Simons EFT Couplings}\label{app:EFT}
The centerpiece of the littlest Higgs model is the $SU(5)/SO(5)$ coset field, whose dynamics are encoded as a NLSM model with Lagrangian
\begin{equation}
\mathcal{L} = \frac{f^2}{8}\mathrm{Tr}\left[\partial_{\mu}\Sigma^{\dagger}\partial^{\mu}\Sigma\right]
\end{equation}
where $\Sigma$ is defined as
\begin{equation}
\Sigma = \exp\left[\frac{2i}{f}\Pi\right]\Braket{\Sigma} = \exp\left[\frac{2i}{f}T_a\pi_a\right]\Braket{\Sigma}
\end{equation}
where $\Pi$ is the `pion' field matrix made by contracting  and $f$ is the `pion' decay constant. This matrix has the following transformation property:
\begin{equation}
\Sigma\rightarrow U\Sigma U^{\mathsf{T}}
\label{eq:Sigma_transform}
\end{equation}
It can be shown that the generators of $SU(5)$ can be classed as either broken or unbroken, $X_a$ and $T_a$, respectively, satisfying these relations
\begin{align*}
T_a\Braket{\Sigma} + \Braket{\Sigma}T_a^{\mathsf{T}} = 0\qquad\text{and}\qquad X_a\Braket{\Sigma} - \Braket{\Sigma}X_a^{\mathsf{T}} = 0
\label{eq:generator_constraints}
\end{align*}
The pion matrix fully written out is
\begin{equation}
\begin{split}
&\Pi = \begin{pmatrix}
-\frac{\omega^0}{2} - \frac{\eta}{\sqrt{20}}	&	-\frac{\omega^+}{\sqrt{2}}	&	\frac{H^+}{\sqrt{2}}	&	-i\phi^{++}	&	-\frac{i\phi^+}{\sqrt{2}}
\\
-\frac{\omega^-}{\sqrt{2}}	&	\frac{\omega^0}{2} - \frac{\eta}{\sqrt{20}}	&	\frac{H^0}{\sqrt{2}}	&	-i\frac{\phi^+}{\sqrt{2}	}&	\frac{-i\phi^0+\phi^0_P}{\sqrt{2}}
\\
\frac{H^-}{\sqrt{2}}	&	\frac{H^{0*}}{\sqrt{2}}		&	\sqrt{\frac{4}{5}}\eta	&	\frac{H^+}{\sqrt{2}}	&	\frac{H^0}{\sqrt{2}}
\\
i\phi^{--}	&	\frac{i\phi^-}{\sqrt{2}}		&	\frac{H^-}{\sqrt{2}}	&	-\frac{\omega^0}{2} - \frac{\eta}{\sqrt{20}}	&	-\frac{\omega^-}{\sqrt{2}}
\\
\frac{i\phi^-}{\sqrt{2}}		&	\frac{i\phi^0+\phi^0_P}{\sqrt{2}}		&	\frac{H^{0*}}{\sqrt{2}}		&	-\frac{\omega^+}{\sqrt{2}}	&	\frac{\omega^0}{2} - \frac{\eta}{\sqrt{20}}
\end{pmatrix}
\\
&= \begin{pmatrix}
\omega - \frac{\eta}{\sqrt{20}}\mathbb{1}_2	&	\frac{H}{\sqrt{2}}	&	\phi^{\dagger}
\\
\frac{H^{\dagger}}{\sqrt{2}}	&	\sqrt{\frac{4}{5}}\eta	&	\frac{H^{\mathsf{T}}}{\sqrt{2}}
\\
\phi	&	\frac{H^*}{\sqrt{2}}	&	\omega^{\mathsf{T}} - \frac{\eta}{\sqrt{20}}\mathbb{1}_2
\end{pmatrix}.
\end{split}
\end{equation}
In addition to the spontaneous breaking induced by $\Braket{\Sigma}$, the $SU(5)$ symmetry is also explicitly broken by gauging an $SU(2)\times [SU(2)\times U(1)]$ subgroup. In the following, $Q^a_{1,2}$ and $Y$ are the generators of the $SU(2)$ and $U(1)$ subgroups of $SU(5)$ that are being gauged and $W^{\mu}_{i,a}$ and $B^{\mu}$ are the gauge fields. With four gauge fields, there are four corresponding couplings: two `weak' couplings $g_1, g_2$ and a hypercharge coupling $g'$. These are generated by
\begin{gather}
Q_1^a = \begin{pmatrix}
\frac{\sigma_a}{2}	&	&
\\
&   &
\\
&   &
\end{pmatrix},
\\
Q_2^a = \begin{pmatrix}
&   &
\\
&   &
\\
&   &	-\frac{\sigma_a^*}{2}
\end{pmatrix},
\\
Y = \frac{1}{2}\begin{pmatrix}
\mathbb{1}_2	&	&
\\
&   & 
\\
&   &
\end{pmatrix}.
\label{eq:subgroup_generators}
\end{gather}
Thus the full covariant derivative is
\begin{equation}
\begin{split}
D_{\mu}\Sigma &= \partial_{\mu}\Sigma
\\
&+ ig_jW^a_{j,\mu}\left(Q^j_a\Sigma + \Sigma Q_a^{j\mathsf{T}}\right)
\\
&+ ig'B_{\mu}\left(Y\Sigma + \Sigma Y\right)
\label{eq:full_covariant_derivative}
\end{split}
\end{equation}
so that the NLSM Lagrangian is
\begin{equation}
\mathcal{L} = \frac{f^2}{8}\mathrm{Tr}\left[\left(D_{\mu}\Sigma\right)^{\dagger}D^{\mu}\Sigma\right].
\end{equation}
The explicit expressions showing the three- and four-point scalar-boson couplings can be found by expanding the $\Sigma$ field
\begin{equation}
\Sigma = \exp\left[\frac{2i}{f}\Pi\right]\Braket{\Sigma} \approx \Braket{\Sigma} + \frac{2i}{f}\Pi\Braket{\Sigma},
\label{eq:expansion}
\end{equation}
where the $\Pi\Braket{\Sigma}$ term is given by
\begin{align}
\Pi\Braket{\Sigma} &=  \begin{pmatrix}
	&	\frac{H}{\sqrt{2}}	&	\phi^{\dagger}
\\
\frac{H^{\dagger}}{\sqrt{2}}	&		&	\frac{H^{\mathsf{T}}}{\sqrt{2}}
\\
\phi	&	\frac{H^*}{\sqrt{2}}	&	
\end{pmatrix}\begin{pmatrix}
	&	&	\mathbb{1}_2
\\
	&	1	&	
\\
\mathbb{1}_2	&		&	
\end{pmatrix} = \begin{pmatrix}
\phi^{\dagger}	&	\frac{H}{\sqrt{2}}	&	
\\
\frac{H^{\mathsf{T}}}{\sqrt{2}}	&		&	\frac{H^{\dagger}}{\sqrt{2}}
\\
	&	\frac{H^*}{\sqrt{2}}	&	\phi
\end{pmatrix}.
\label{eq:pion_vac_product}
\end{align}
We can leverage what we've learned thus far to engender a coupling between the scalar sector and the Chern-Simons term. The unique gauge-invariant term that is second order in $\Sigma$ and yields a tree-level quadratic interaction is of the form
\begin{equation}
\text{Tr}\left[\left(Y\Sigma\right)\left(Y\Sigma\right)^*\right].
\end{equation}
This term is previously used when examining the gauge boson mass matrix;  in turn may be multiplied by $\text{Tr}[W\tilde{W}]$, which is itself already gauge invariant:
\begin{equation}
\sum_{j = 1}^{2}\frac{g_j^2 \beta^2 }{16}\text{Tr}\left[\left(Y\Sigma\right)\left(Y\Sigma\right)^*\right]\text{Tr}\left[\tilde{W}_j^a W^j_a\right].
\end{equation}
Here $\beta$ is a constant determined by the UV complete theory, to which we are agnostic. Expanding the trace of $\Sigma$ fields to lowest order in $\Pi$ yields a mass term for the Higgs doublet proportional to $g'$, the $U(1)_Y$ subgroup of $SU(5)$. This is to be expected, as the gauged subgroup explicitly breaks $SU(5)$ and the shift symmetry prohibiting mass terms with it. As shown in deriving Eq.~\eqref{eq:BSLHP}, a quadratic Higgs interaction of the form
\begin{align}
 &\sum_{j=1}^2\frac{g^2_j\beta^2}{16}\text{Tr}\left[\left(Y\Sigma\right)\left(Y\Sigma\right)^*\right]\text{Tr}\left[\tilde{W}_jW_j\right]
 \\
 &\rightarrow \frac{\beta^2}{2f^2}\left(H^{\dagger}H\right)\sum_{j=1}^2g_j^2\text{Tr}\left[\tilde{W}W\right].
\end{align}
arises, coupling to the CS current, as desired.

\section{Electron Electric Dipole Moment}\label{app:eEDM}
We calculate the induced electric dipole moment (EDM) of the electron from the new Chern-Simons operator. Below EWSB, the Higgs doublet is parameterized as $H = [0, (v + h)/\sqrt{2}]^\mathsf{T}$, with $v = 246\ {\rm GeV}$ the Higgs electroweak VEV and $h$ now representing fluctuations from this value (not to be confused with the background $h$ elsewhere in the text). Moreover, the hypercharge gauge field $B_\mu$ and the third weak-isospin gauge boson $W_\mu^3$ mix to form the massive $Z$-boson $Z_\mu$ and massless photon $A_\mu$. In particular, the third weak-isospin gauge boson is given by the linear combination $W_\mu^3 = \sin\left(\theta_{\rm W}\right)A_\mu + \cos\left(\theta_{\rm W}\right)Z_\mu$, giving rise to the electromagnetic Chern-Simons operator
\begin{align}\label{eq:hFF}
\frac{g^2 \beta^2}{2f^2}\left(H^\dagger H\right)\left[W_a^{\mu\nu}\tilde{W}_{\mu\nu}^a\right] \supset \frac{e^2\beta^2}{4 f^2}\left(h v\right)F^{\mu\nu}\tilde{F}_{\mu\nu},
\end{align}
where $F_{\mu\nu} = \partial_\mu A_\nu - \partial_\nu A_\mu$ is the electromagnetic field strength tensor and we used the gauge-coupling relation $g  = e/\sin\left(\theta_{\rm W}\right)$. Together with the SM Yukawa electron-Higgs interaction,
\begin{align}
y_1 L_1 H e_R \supset \frac{y_1}{\sqrt{2}} h \bar{\Psi}_e \Psi_e,
\end{align}
 as well as the covariant-derivative electron-photon interaction
\begin{align}
i\bar{\Psi}_e\gamma^\mu D_\mu \Psi_e \supset e A_\mu \bar{\Psi}_e\gamma^\mu \Psi_e,
\end{align}
the electromagnetic Chern-Simons operator induces an electron EDM through the triangle diagram in Fig.~\ref{fig:eEDM}. In the above expressions, $L_1 = \left(\nu_{e L}, e_L\right)$ is the first generation lepton doublet,  $\Psi_e = \left(e_L, e_R\right)$ is the electron Dirac spinor, and $e_i$, $i \in \{L, R\}$,  are the corresponding  electron chirality states. 

The induced electron electric dipole moment is then~\cite{Lue:1996pr}
\begin{align}
\frac{d_e}{e} &= \frac{m_e \sin^2\left(\theta_{\rm W}\right)}{8\pi^2}\frac{g^2 \beta^2}{2 f^2}\log\left(\frac{f^2 + m_H^2}{m_H^2}\right).
\end{align}

\nocite*
\bibliographystyle{hunsrt}
\bibliography{pNGB_higgs}

\end{document}